\documentclass[epjST]{svjour}
\usepackage{graphicx}
\usepackage{amsmath}
\usepackage{mathtools}
\usepackage{amssymb}
\usepackage{bm}
\usepackage{pifont}
\usepackage{color}
\usepackage[FIGTOPCAP]{subfigure}
\usepackage[latin1]{inputenc}

\begin{document}
%
%
%
%

\title{Discussion on Peshkov et al., ``Boltzmann-Ginzburg-Landau approach 
for continuous descriptions of generic Vicsek-like models''}

\author{Thomas Ihle\inst{1}\fnmsep\thanks{\email{thomas.ihle@ndsu.edu}}} 

\institute{Department of Physics, North Dakota State University, Fargo, ND 58108-6050, USA}

\abstract{
A discussion on the contribution of Peshkov, Bertin, Ginelli and Chat{\'e} \cite{peshkov_ST} in this special issue.
}

\maketitle


In contribution \cite{peshkov_ST}, the authors review a theoretical approach, called BGL, that is based on a 
Boltzmann equation 
for point-like active particles. The particles are assumed to interact through alignment interactions 
similar to the 
ones of the Vicsek-model \cite{vicsek_95}.
The aim of their approach is to describe
the large scale behavior of active matter.
Because this goal is identical to the one of the phase space approach (PSA), also called Enskog equation approach, presented in this special issue \cite{ihle_ST},
I first would like to comment on the general differences between the two approaches.
First, BGL and PSA seem to have opposing research philosophies.
As I see it, Peshkov {\em et al.} try to set up a qualitative description, as simple as possible, by relying on a 
concept 
from critical phenomena that probably there is some type of universality and that only a few terms in the  
hydrodynamic equations 
will suffice to understand the behavior at large scales.
Another argument to justify their minimalistic approach appears 
to be that the underlying microscopic models like the Vicsek-model are often quite 
unrealistic, so why bother with sticking too close to these models.
My approach is quite the opposite: 
I take microscopic models at face value and try to set up a theory that is quantitative and describes the underlying model
as close as possible. The hope is, that if one manages to fully understand a particular model, even if it is simplistic,
one might get ideas about how to deal with more realistic situations. 
Moreover, by considering quantitative theories for
similar but microscopically different models
one could test what aspects of the models actually are universal.

I do not want to diminish the insights gained by the BGL approach on the qualitative level
but I do believe it is important to strive for theories that are built on solid ground. 
My main criticism of BGL is that its foundation appears shaky.
This is because BGL's hydrodynamic equations are derived from the Boltzmann equation whose validity
is based on two main assumptions. The first one assumes binary collions but 
as shown in \cite{ihle_ST}, even in the dilute limit this is not valid in Vicsek-like models at realistic 
velocities and time steps, 
in and close to the state of collective motion.
The second assumption -- Molecular Chaos -- is also not valid in such dilute active particle systems close to the transition, see 
Refs. \cite{chou_12,hanke_13,chou_14}, unless particles are allowed to be invisible to each other for sufficiently long times.
Thus, while I do not doubt that the authors derived their hydrodynamic equations in a controlled way 
(within the inherent limitations of an asymptotic expansion) from
the Boltzmann equation, their starting point seems questionable.
The derived macroscopic equations might still work, 
at least qualitatively, but cannot be completely trusted. 
There could even be cases where qualitative phenomena are missed. For example, to my knowledge,
BGL has not yet managaged to reproduce the scaling and massive system size dependence of solitary waves 
that were found by both the PSA-approach and agent-based simulations 
\cite{ihle_13}.

I see the PSA approach \cite{ihle_ST,ihle_11} as well as the cluster approach of Refs. \cite{peruani_06,peruani_13} 
as first steps to remedify the shortcomings of a Boltzmann approach by including, at least approximately, the ability to 
handle dense systems with non-binary interactions and pre-collisional correlations.
Currently, PSA does apply at arbitrary number density and 
controls 
the Molecular Chaos (MC) approximation 
by choosing a large enough mean free path (mfp).
An extension which works at more realistic smaller mfp, includes correlations and goes beyond MC, will be published elsewhere \cite{chou_14}.

Let me now comment on specific parts of the contribution by Peshkov {\em et al.}.
In Section C2, Eq. (40), the authors define the scaling relations for the gradients $\nabla\sim \epsilon$ and Fourier coefficients
$f_k\sim \epsilon^k$ for $k>0$, using a formal parameter $\epsilon$. 
An expansion of the Boltzmann equation in powers of $\epsilon$ then yields the
hydrodynamic equation, Eq. (58).
The above scaling assumptions are exactly the same as the ones used in PSA, \cite{ihle_ST} but there 
is a subtle difference in the scaling of the density, which can lead to misunderstandings.
Peshkov {\em et al.} use $\rho-\rho_0\sim\epsilon$ (polar model with ferromagnetic alignment)
that translates into the relation $f_0=\rho_0/(2\pi)+O(\epsilon)$ for the $k=0$ Fourier coefficient.
That means, $f_0$ is a sum of a constant of order one and a term of order $\epsilon$, and for example, leads to the scaling, $\nabla \rho\sim \epsilon^2$. In contrast, the Chapman-Enskog exansion of Refs. \cite{ihle_ST,ihle_11} assumes
$f_0\sim O(1)$, leading to the scaling $\nabla \rho\sim \epsilon$.
Therefore, on one hand, from PSA's scaling perspective, Eq. (58) looks as if it misses many contributions of order $O(\epsilon^3)$.
On the other hand, analysing the hydrodynamic equations of PSA by means of BGL's scaling laws gives the impression
that it contains too many density gradients up to order $O(\epsilon^6)$ but omits momentum density terms of the same order.

In principle, the properties of the collision operator determine which scaling choices are admissable. 
Since we are interested in the large scale behavior it also makes sense to dress gradients with some power of $\epsilon$. 
However, there seems to be
some ambiguity in these choices. As
long as no inconsistencies occur in the respective gradient or Chapman-Enskog expansions, 
ultimately 
the same terms will be recovered, just at different orders of the expansion and sometimes broken into several pieces. 
Let's also keep in mind, that $\epsilon$ is just a formal
expansion parameter 
which is set to one at the end of the derivation. 
The decision about which term is smaller than another, e.g. carries a higher power of 
$\epsilon$ is often based on balancing certain terms \cite{peshkov_ST} and on the 
spatio-temporal behavior of small deviations from a simple stationary solution.
For BGL, the simplest stationary state, the homogeneous disordered state, is chosen.  
While a certain ``balancing'' might be adequate close to this reference state, there is no guarantee that it still works,
once the system starts forming strongly nonlinear structures such as steep density waves.
I believe it is worthwhile to look closer into the ambiguity of scaling choices and to use 
the numerical procedures of Refs. \cite{ihle_13,thueroff_13} in order to judge which nonlinearities are relevant in steep 
waves.
I agree with the statement of the authors below Eq. (40) that 
not all important terms need to be balanced and that 
a multiscale expansion would be useful. 
I would like to point point out that such an expansion was 
part of the Chapman-Enskog procedure for PSA, \cite{ihle_ST,ihle_11}, where one fast and three different slow 
time scales were formally 
included. 

Around Eqs. (74)-(83), the authors discuss the derivation of hydrodynamic equations 
and find that the equations become more and more unreliable, once they increase the order of the expansion beyond three.
While this result agrees with my prediction \cite{ihle_ST} I am more optimistic than their conclusion ``\ldots
an angular Fourier expansion is not suited far from the onset of order ..''
While it is true that at the extreme limit of zero noise, all Fourier coefficients $f_k$ for $k>0$ take on the same value and 
none can be neglected, 
in my experience these expansions are fairly well behaved as long as no gradient expansions are performed and the noise is not 
too small. 
For example, an expansion that keeps the time evolution of
the first modes up to $k=5$, produced nice results even for strongly inhomogeneous solutions \cite{ihle_13}.
Therefore, I rather expect that not the Fourier coefficients are the problem but the gradient expansion.

In Section IV.C., the authors write that ``propagating bands'' in the polar case were explained in Caussin {\em et al.} \cite{caussin_14}.
I would like to mention that for the Vicsek-model, 
it was already explained in Ref. \cite{ihle_13} using kinetic theory, how these solitary bands behave and
provide a mean-field mechanism to render the order/disorder transition discontinuous.

Finally, in the introduction and conclusion, the authors refer to the 
classification of dry active matter in the ``polar class'',
``active nematics'' and polar particles aligning nematically, called ``self-propelled rods''. 
This is a good initial classification but I think one has to be cautious to not overstretch the implied concept of universality. 
For example, according to this categorization, 
both the regular Vicsek-model \cite{vicsek_95} and its metric-free version
are in the ``polar class'' and their hydrodynamic equations are supposed to have the same structure \cite{chou_12,peshkov_12}.
However, the nature of the transition to collective motion is still different: 
discontinuous for the regular VM but continuous for the metric-free model.
Another example are real self-propelled rods with excluded volume interactions \cite{peruani_06} and the 
Vicsek-model with nematic alignment \cite{peruani_08,ginelli_10}. 
Both are in the ``self-propelled rod''-class, but in the latter, a phase is possible that consists of a dense, immobile
band of nematic order, whereas real rods cannot have such a phase but rather show flocks of polar order \cite{peruani_14}.
In general, excluded volume effects or other mechanisms that introduce a coupling between particle speed and density
\cite{ohta_ST,peruani_11} can lead to a zoology of complex patterns, and one might have to refine the classification of active matter.

\end{document}